\documentclass[12pt]{article}%
\usepackage{amsmath}
\usepackage{amsfonts}
\usepackage{amssymb}
\usepackage{graphicx}%
\newtheorem{theorem}{Theorem}

\newtheorem{proposition}[theorem]{Proposition}

\begin{document}

\title{Higher Toda Mechanics and Spectral Curves}
\author{Liu Zhao\thanks{e-mail: lzhao@nwu.edu.cn}, Wangyun Liu\thanks{e-mail:
wyl@phy.nwu.edu.cn}\\Institute of Modern Physics,\\Northwest University, Xian 710069, China}
\maketitle

\begin{abstract}
For each one of the Lie algebras $\mathfrak{gl}_{n}$ and $\widetilde
{\mathfrak{gl}}_{n}$, we constructed a family of integrable generalizations of
the Toda chains characterized by two integers $m_{+}$ and $m_{-}$. The Lax
matrices and the equations of motion are given explicitly, and the integrals
of motion can be calculated in terms of the trace of powers of the Lax matrix
$L$. For the case of $m_{+}=m_{-}$, we find a symmetric reduction for each
generalized Toda chain we found, and the solution to the initial value
problems of the reduced systems is outlined. We also studied the spectral
curves of the periodic $(m_{+},m_{-})$-Toda chains, which turns out to be very
different for different pairs of $m_{+}$ and $m_{-}$. Finally we also obtained
the nonabelian generalizations of the $(m_{+},m_{-})$-Toda chains in explicit form.

\end{abstract}

\section{Introduction}

The very first research interests in classical many body systems can be traced
back to the dates from Galileo to Newton, when the classical mechanics for the
celestial bodies in the solar system was the major physical system of concern.
However, it turns out that it is extremely difficult, or even impossible in
most cases, to get exact solutions to the equations of motion for a given many
body mechanical system. Perhaps the major break through in this direction was
made since the late 1960's and the early 1970's, when a number of exactly
solvable many body systems in one dimension were discovered and solved by
means of the inverse scattering method, among which the classical Toda chains
\cite{1,2}, the Calogero-Moser systems \cite{3,4} and the
Ruijsenaars-Schneider models \cite{8,7} are the most famous and important examples.

The recent surge of interests in these classical mechanical systems is mostly
due to the work of Donagi and Witten \cite{5} and the followers \cite{6},
which revealed a remarkable and unexpected relationship between such solvable
mechanical systems and certain $N=2$ supersymmetric gauge theories. To be more
specific, the exact integrability of the above many body systems are
characterized by their Lax matrices, whose spectral curves in turn are
identified with the so-called Seiberg-Witten curves  \cite{9} of the $N=2$
SUSY gauge theories, while the complete information about the moduli space
parameters and the prepotentials of the $N=2$ SUSY gauge theories are encoded
in the equation defining the Seiberg-Witten curves.

More recently, Dijkgraaf and Vafa made another important progress \cite{DV} in
finding nonperturbative information about the moduli structure of SUSY gauge
theories. They find a remarkable correspondence between the nonperturbative
chiral effective superpotentials and the effective potentials of certain
matrix integrals in the planner diagram limit. Their results fully conformed
the earlier results of Dorey \cite{Dorey}, who identified the same chiral
superpotentials with the values of the Hamiltonians of certain integrable
classical many body systems in the stationary point configuration. Although
this last step toward the nonperturbative treatment of $N=1$ SUSY gauge
theories is mathematically less complete compared to the corresponding
Seiberg-Witten theory for $N=2$ cases, it has caught quite massive attention
because the $N=1$ SUSY theories are considered to be more realistic than the
$N=2$ ones.

So far, the classical many body systems known to play certain role in the
studies in both the $N=2$ and $N=1$ gauge theories is restricted within the
above mentioned three families, i.e. the classical Toda chains, the
Calogero-Moser systems and the Ruijsenaars-Schneider models. It seems an
interesting question to ask whether there exist other pairs of SUSY gauge
theory and classical many body systems and whether the known classes of SUSY
gauge theories may be related to other many body systems besides the ones just
listed and vice versa. In this paper we shall try to construct some novel
exactly solvable many body systems which fall in the family of Toda mechanics,
but with off-diagonal extensions (the meaning of off-diagonal extension will
be clear in the main context), and study their spectral curves. We hope some
day such systems will be proved useful in describing certain SUSY gauge theory
in a nonperturbative way, following either the Seiberg-Witten or the
Dijkgraaf-Vafa routes. But now we shall mostly concentrate on the integrable
aspect of these extended Toda mechanical systems.

This paper is organized as follows. In Section 2 we give the Lax matrix
construction of the extended Toda mechanics, referred to as $(m_{+},m_{-}%
)$-Toda chains. The construction applies for infinite chains, finite chains as
well as periodic chains. The Liouville integrability of the finite
$(m_{+},m_{-})$-Toda chains is discussed via the construction for integrals of
motion and the definition for the fundamental Poisson structure. For the case
of $m_{+}=m_{-}$, we show that there is a symmetric reduction to the
$(m_{+},m_{-})$-Toda chains and the solution to their initial value problems
are described. Section 3 is devoted to a brief introduction for the spectral
curves corresponding to the extended periodic chains. It is shown that there
are two different ways for describing the spectral curves, each has a
correspondence in the cases of standard Toda chains. In Section 4, we
generalize our construction for the extended Toda chains to the nonabelian
cases and explicit equations of motion for generic nonabelian $(m_{+},m_{-}%
)$-Toda chains are obtained. The paper is then concluded in Section 5, where
we present some discussions and speculations for further studies.

\section{Construction of the system}

\bigskip In this section we shall mainly work on the construction of the
extended Toda mechanics. We shall follow the route of Lax integrability during
the construction, give necessary explanation for the construction and briefly
discuss the Liouville integrability of the systems. The systems we shall be
considering can be divided into three classes: the infinite chains, the finite
chains and the periodic chains. It is the periodic chains which have
nontrivial spectral curves. We leave the detailed study on the spectral curve
to the next section.

\subsection{Infinite chain}

Let us first introduce some auxiliary notations. We shall need an infinite
dimensional vector space with basis vectors $\{|i\rangle,i\in\mathbb{Z}\}$ and
their duals, $\{\langle i|,i\in\mathbb{Z}\}$. The inner product between the
basis vectors and their duals is specified by%
\[
\langle i|j\rangle=\delta_{ij}.
\]
Let us remind that such vectors are introduced purely for notational
conveniences, they have nothing to do with the Fock space basis vectors for
the Heisenberg algebras.

With the above notations in mind, we are now ready to introduce the Lax
matrices for the infinite chain of extended Toda mechanics. By Lax matrices we
mean the pair $(\mathcal{L},\mathcal{M})$ of matrices appearing in the Lax
equation
\begin{equation}
\mathcal{\dot{L}}=[\mathcal{M},\mathcal{L}], \label{Lax}%
\end{equation}
where, as usual, a dot over the mechanical variables means taking a time
derivative. For the case of infinite chains we shall introduce, the matrix
$\mathcal{L}$ is a formal infinite matrix with a strip form, i.e. all nonzero
entries of $\mathcal{L}$ are at positions close to the main diagonal. More
concretely, $\mathcal{L}$ may be written as follows,%
\begin{align}
\mathcal{L}  &  =\mathcal{L}^{(0)}+\mathcal{L}^{(>)}+\mathcal{L}%
^{(<)},\label{L}\\
\mathcal{L}^{(0)}  &  =%
%TCIMACRO{\dsum \limits_{i\in\mathbb{Z}}}%
%BeginExpansion
{\displaystyle\sum\limits_{i\in\mathbb{Z}}}
%EndExpansion
\dot{q}_{i}|i\rangle\langle i|,\label{L1}\\
\mathcal{L}^{(>)}  &  =%
%TCIMACRO{\dsum \limits_{i\in\mathbb{Z}}}%
%BeginExpansion
{\displaystyle\sum\limits_{i\in\mathbb{Z}}}
%EndExpansion%
%TCIMACRO{\dsum \limits_{k=1}^{m_{+}}}%
%BeginExpansion
{\displaystyle\sum\limits_{k=1}^{m_{+}}}
%EndExpansion
a_{i}^{(k)}|i\rangle\langle i+k|,\label{L2}\\
\mathcal{L}^{(<)}  &  =%
%TCIMACRO{\dsum \limits_{i\in\mathbb{Z}}}%
%BeginExpansion
{\displaystyle\sum\limits_{i\in\mathbb{Z}}}
%EndExpansion%
%TCIMACRO{\dsum \limits_{k=1}^{m_{-}}}%
%BeginExpansion
{\displaystyle\sum\limits_{k=1}^{m_{-}}}
%EndExpansion
c_{i}^{(k)}|i+k\rangle\langle i|. \label{L3}%
\end{align}
Each $\mathcal{L}$ matrix is characterized with a pair of fixed integers
$(m_{+},m_{-})$, with $m_{+}+m_{-}+1$ being the width of the nonzero strip in
the $\mathcal{L}$ matrix. We shall see that for $m_{+}=m_{-}=1$, the matrix
$\mathcal{L}$ would yield the standard infinite Toda chain. Therefore we may
call the system with generic values of $m_{\pm}$ an $(m_{+},m_{-})$-th
extended Toda chain. In what follows, we shall always assume that $m_{+}\geq
m_{-}$, without loss of generality.

Now let us proceed to specify the $\mathcal{M}$ matrix which is needed to
obtain the equation of motion for a Lax integrable system via the Lax equation
(\ref{Lax}). The $\mathcal{M}$ matrix for the present case is simply given as%
\begin{equation}
\mathcal{M}=\mathcal{L}^{(>)}-\mathcal{L}^{(<)}. \label{M}%
\end{equation}
Therefore, the right hand side of the Lax equation may be written as%
\[
\lbrack\mathcal{M},\mathcal{L}]=[\mathcal{L}^{(>)}-\mathcal{L}^{(<)}%
,\mathcal{L}^{(0)}]+2[\mathcal{L}^{(>)},\mathcal{L}^{(<)}].
\]
Straightforward calculations yield the following results,%
\begin{align*}
\lbrack\mathcal{L}^{(>)},\mathcal{L}^{(0)}]  &  =%
%TCIMACRO{\dsum \limits_{i\in\mathbb{Z}}}%
%BeginExpansion
{\displaystyle\sum\limits_{i\in\mathbb{Z}}}
%EndExpansion%
%TCIMACRO{\dsum \limits_{k=1}^{m_{+}}}%
%BeginExpansion
{\displaystyle\sum\limits_{k=1}^{m_{+}}}
%EndExpansion
(a_{i}^{(k)}\dot{q}_{i+k}-\dot{q}_{i}a_{i}^{(k)})|i\rangle\langle i+k|,\\
\lbrack\mathcal{L}^{(<)},\mathcal{L}^{(0)}]  &  =-%
%TCIMACRO{\dsum \limits_{i\in\mathbb{Z}}}%
%BeginExpansion
{\displaystyle\sum\limits_{i\in\mathbb{Z}}}
%EndExpansion%
%TCIMACRO{\dsum \limits_{k=1}^{m_{-}}}%
%BeginExpansion
{\displaystyle\sum\limits_{k=1}^{m_{-}}}
%EndExpansion
(\dot{q}_{i+k}c_{i}^{(k)}-c_{i}^{(k)}\dot{q}_{i})|i+k\rangle\langle i|,
\end{align*}
\begin{align*}
\lbrack\mathcal{L}^{(>)},\mathcal{L}^{(<)}]  &  =%
%TCIMACRO{\dsum \limits_{i\in\mathbb{Z}}}%
%BeginExpansion
{\displaystyle\sum\limits_{i\in\mathbb{Z}}}
%EndExpansion%
%TCIMACRO{\dsum \limits_{k=1}^{m_{-}}}%
%BeginExpansion
{\displaystyle\sum\limits_{k=1}^{m_{-}}}
%EndExpansion
\left\{  a_{i}^{(k)}c_{i}^{(k)}-c_{i-k}^{(k)}a_{i-k}^{(k)}\right\}
|i\rangle\langle i|\\
&  +%
%TCIMACRO{\dsum \limits_{i\in\mathbb{Z}}}%
%BeginExpansion
{\displaystyle\sum\limits_{i\in\mathbb{Z}}}
%EndExpansion%
%TCIMACRO{\dsum \limits_{k=1}^{m_{+}-1}}%
%BeginExpansion
{\displaystyle\sum\limits_{k=1}^{m_{+}-1}}
%EndExpansion
\left(
%TCIMACRO{\dsum \limits_{k^{\prime}=1}^{\min(m_{+}-k,m_{-})}}%
%BeginExpansion
{\displaystyle\sum\limits_{k^{\prime}=1}^{\min(m_{+}-k,m_{-})}}
%EndExpansion
\left(  a_{i}^{(k+k^{\prime})}c_{i+k}^{(k^{\prime})}-c_{i-k^{\prime}%
}^{(k^{\prime})}a_{i-k^{\prime}}^{(k+k^{\prime})}\right)  \right)
|i\rangle\langle i+k|\\
&  +%
%TCIMACRO{\dsum \limits_{i\in\mathbb{Z}}}%
%BeginExpansion
{\displaystyle\sum\limits_{i\in\mathbb{Z}}}
%EndExpansion%
%TCIMACRO{\dsum \limits_{k=1}^{m_{-}-1}}%
%BeginExpansion
{\displaystyle\sum\limits_{k=1}^{m_{-}-1}}
%EndExpansion
\left(
%TCIMACRO{\dsum \limits_{k^{\prime}=1}^{m_{-}-k}}%
%BeginExpansion
{\displaystyle\sum\limits_{k^{\prime}=1}^{m_{-}-k}}
%EndExpansion
\left(  a_{i+k}^{(k^{\prime})}c_{i}^{(k+k^{\prime})}-c_{i-k^{\prime}%
}^{(k+k^{\prime})}a_{i-k^{\prime}}^{(k^{\prime})}\right)  \right)
|i+k\rangle\langle i|.
\end{align*}
Therefore, the equations of motion obtained from the Lax equation read%
\begin{align}
\ddot{q}_{i}  &  =2%
%TCIMACRO{\dsum \limits_{k=1}^{m_{-}}}%
%BeginExpansion
{\displaystyle\sum\limits_{k=1}^{m_{-}}}
%EndExpansion
\left\{  a_{i}^{(k)}c_{i}^{(k)}-c_{i-k}^{(k)}a_{i-k}^{(k)}\right\}
,\label{eq:1}\\
\dot{a}_{i}^{(k)}  &  =a_{i}^{(k)}\dot{q}_{i+k}-\dot{q}_{i}a_{i}%
^{(k)}\nonumber\\
&  +2%
%TCIMACRO{\dsum \limits_{k^{\prime}=1}^{\min(m_{+}-k,m_{-})}}%
%BeginExpansion
{\displaystyle\sum\limits_{k^{\prime}=1}^{\min(m_{+}-k,m_{-})}}
%EndExpansion
\left(  a_{i}^{(k+k^{\prime})}c_{i+k}^{(k^{\prime})}-c_{i-k^{\prime}%
}^{(k^{\prime})}a_{i-k^{\prime}}^{(k+k^{\prime})}\right)  ,\qquad(1\leq k\leq
m_{+}-1)\label{eq:2}\\
\dot{c}_{i}^{(k)}  &  =\dot{q}_{i+k}c_{i}^{(k)}-c_{i}^{(k)}\dot{q}%
_{i}\nonumber\\
&  +2%
%TCIMACRO{\dsum \limits_{k^{\prime}=1}^{m_{-}-k}}%
%BeginExpansion
{\displaystyle\sum\limits_{k^{\prime}=1}^{m_{-}-k}}
%EndExpansion
\left(  a_{i+k}^{(k^{\prime})}c_{i}^{(k+k^{\prime})}-c_{i-k^{\prime}%
}^{(k+k^{\prime})}a_{i-k^{\prime}}^{(k^{\prime})}\right)  ,\qquad(1\leq k\leq
m_{-}-1)\label{eq:3}\\
\dot{a}_{i}^{(m_{+})}  &  =a_{i}^{(m_{+})}\dot{q}_{i+m_{+}}-\dot{q}_{i}%
a_{i}^{(m_{+})},\label{eq:4}\\
\dot{c}_{i}^{(m_{-})}  &  =\dot{q}_{i+m_{-}}c_{i}^{(m_{-})}-c_{i}^{(m_{-}%
)}\dot{q}_{i}. \label{eq:5}%
\end{align}
Notice that while writing the above equations we have been very careful in
keeping the order of the variables in their essential positions, with the hope
of generalizing to the case of nonabelian variables in mind. However, at
present, let us proceed with the view point that all variables $q_{i}%
,a_{i}^{(k)}$ and $c_{i}^{(k)}$ are usual abelian variables. Then it is easy
to see that the last pair of the equations of motion, eqs.(\ref{eq:4}),
(\ref{eq:5}) can be integrated explicitly, yielding%
\begin{equation}
a_{i}^{(m_{+})}=\exp(q_{i+m_{+}}-q_{i}),\quad c_{i}^{(m_{-})}=\exp(q_{i+m_{-}%
}-q_{i}). \label{am}%
\end{equation}
Similarly, we can rewrite the variables $a_{i}^{(k)}$ $(1\leq k<m_{+})$ and
$c_{i}^{(k)}$ $(1\leq k<m_{-})$ in the form%
\begin{equation}
a_{i}^{(k)}=\exp(q_{i+k}-q_{i})\psi_{i}^{+(k)},\quad c_{i}^{(k)}=\exp
(q_{i+k}-q_{i})\psi_{i}^{-(k)}, \label{subs}%
\end{equation}
so that the equations (\ref{eq:1}-\ref{eq:3}) are turned into the following
form in terms of the new variables $\psi_{i}^{\pm(k)}$,%
\begin{align}
\ddot{q}_{i}  &  =2%
%TCIMACRO{\dsum \limits_{k=1}^{m_{-}}}%
%BeginExpansion
{\displaystyle\sum\limits_{k=1}^{m_{-}}}
%EndExpansion
\left\{  \psi_{i}^{+(k)}\psi_{i}^{-(k)}\exp\left[  2(q_{i+k}-q_{i})\right]
-\psi_{i-k}^{-(k)}\psi_{i-k}^{+(k)}\exp\left[  2(q_{i}-q_{i-k})\right]
\right\}  ,\label{nonsym:1}\\
\dot{\psi}_{i}^{+(k)}  &  =2%
%TCIMACRO{\dsum \limits_{k^{\prime}=1}^{\min(m_{+}-k,m_{-})}}%
%BeginExpansion
{\displaystyle\sum\limits_{k^{\prime}=1}^{\min(m_{+}-k,m_{-})}}
%EndExpansion
\left(  \psi_{i}^{+(k+k^{\prime})}\psi_{i+k}^{-(k^{\prime})}\exp\left[
2(q_{i+k+k^{\prime}}-q_{i+k})\right]  \right. \nonumber\\
&  \left.  -\psi_{i-k^{\prime}}^{-(k^{\prime})}\psi_{i-k^{\prime}%
}^{+(k+k^{\prime})}\exp\left[  2(q_{i}-q_{i-k^{\prime}})\right]  \right)
,\quad(1\leq k\leq m_{+}-1)\label{nonsym:2}\\
\dot{\psi}_{i}^{-(k)}  &  =2%
%TCIMACRO{\dsum \limits_{k^{\prime}=1}^{m_{-}-k}}%
%BeginExpansion
{\displaystyle\sum\limits_{k^{\prime}=1}^{m_{-}-k}}
%EndExpansion
\left(  \psi_{i+k}^{+(k^{\prime})}\psi_{i}^{-(k+k^{\prime})}\exp\left[
2(q_{i+k+k^{\prime}}-q_{i+k})\right]  \right. \nonumber\\
&  \left.  -\psi_{i-k^{\prime}}^{-(k+k^{\prime})}\psi_{i-k^{\prime}%
}^{+(k^{\prime})}\exp\left[  2(q_{i}-q_{i-k^{\prime}})\right]  \right)
,\quad(1\leq k\leq m_{-}-1). \label{nonsym:3}%
\end{align}
In the last equations, we have supplemented the substitution rule (\ref{subs})
with $\psi_{i}^{\pm(m_{\pm})}=1$ according to (\ref{am}). The coupled system
of equations (\ref{nonsym:1})-(\ref{nonsym:3}) constitute the set of equations
of motion for our $(m_{+},m_{-})$-th extended Toda chain.

Among all possible pairs of integers $(m_{+},m_{-})$, the special choices
$m_{+}=m_{-}=m$ have some special features. In these cases, the equations of
motion for the $(m,m)$-th extended Toda chain is completely symmetric under
$\psi_{i}^{+(k)}\leftrightarrow\psi_{i}^{-(k)}$ and thus admit a
\emph{symmetric reduction }$\psi_{i}^{+(k)}=\psi_{i}^{-(k)}=\psi_{i}^{(k)}$.
The reduced system of equations of motion reads%
\begin{align}
\ddot{q}_{i}  &  =2%
%TCIMACRO{\dsum \limits_{k=1}^{m-1}}%
%BeginExpansion
{\displaystyle\sum\limits_{k=1}^{m-1}}
%EndExpansion
\left\{  \left(  \psi_{i}^{(k)}\right)  ^{2}\exp\left[  2(q_{i+k}%
-q_{i})\right]  -\left(  \psi_{i-k}^{(k)}\right)  ^{2}\exp\left[
2(q_{i}-q_{i-k})\right]  \right\} \nonumber\\
&  +2\left\{  \exp\left[  2(q_{i+m}-q_{i})\right]  -\exp\left[  2(q_{i}%
-q_{i-m})\right]  \right\}  ,\label{NewEQ2}\\
\dot{\psi}_{i}^{(k)}  &  =2%
%TCIMACRO{\dsum \limits_{k^{\prime}=1}^{m-k}}%
%BeginExpansion
{\displaystyle\sum\limits_{k^{\prime}=1}^{m-k}}
%EndExpansion
\left(  \psi_{i}^{(k+k^{\prime})}\psi_{i+k}^{(k^{\prime})}\exp\left[
2(q_{i+k+k^{\prime}}-q_{i+k})\right]  -\psi_{i-k^{\prime}}^{(k+k^{\prime}%
)}\psi_{i-k^{\prime}}^{(k^{\prime})}\exp\left[  2(q_{i}-q_{i-k^{\prime}%
})\right]  \right)  ,\nonumber\\
(1  &  \leq k\leq m-1). \label{NewEQ1}%
\end{align}
In particular, the special case of $m=1$ gives rise to the standard infinite
Toda chain.

To give the readers some more intuitive idea about the structure for the
$(m_{+},m_{-})$-th extended Toda chain, let us write explicitly several more
cases with concrete values for $m_{+}$ and $m_{-}$, i.e. the cases for
$m_{+},m_{-}\leq3$. These include, beside the simplest case of $m_{+}=m_{-}=1$
(which is automatically a symmetric reduced case, i.e. the standard Toda
chain), the following cases:

\begin{itemize}
\item $m_{+}=2,$ $m_{-}=1$. The corresponding equations of motion read%
\begin{align*}
\ddot{q}_{i}  &  =2\left\{  \psi_{i}^{+(1)}\exp\left[  2(q_{i+1}%
-q_{i})\right]  -\psi_{i-1}^{+(1)}\exp\left[  2(q_{i}-q_{i-1})\right]
\right\}  ,\\
\dot{\psi}_{i}^{+(1)}  &  =2\left\{  \exp\left[  2(q_{i+2}-q_{i+1})\right]
-\exp\left[  2(q_{i}-q_{i-1})\right]  \right\}  ;
\end{align*}

\item $m_{+}=2,$ $m_{-}=2$. The corresponding equations of motion read%
\begin{align*}
\ddot{q}_{i}  &  =2\left\{  \psi_{i}^{+(1)}\psi_{i}^{-(1)}\exp\left[
2(q_{i+1}-q_{i})\right]  -\psi_{i-1}^{-(1)}\psi_{i-1}^{+(1)}\exp\left[
2(q_{i}-q_{i-1})\right]  \right\} \\
&  +2\left\{  \exp\left[  2(q_{i+2}-q_{i})\right]  -\exp\left[  2(q_{i}%
-q_{i-2})\right]  \right\}  ,\\
\dot{\psi}_{i}^{+(1)}  &  =2\left\{  \psi_{i+1}^{-(1)}\exp\left[
2(q_{i+2}-q_{i+1})\right]  -\psi_{i-1}^{-(1)}\exp\left[  2(q_{i}%
-q_{i-1})\right]  \right\}  ,\\
\dot{\psi}_{i}^{-(1)}  &  =2\left\{  \psi_{i+1}^{+(1)}\exp\left[
2(q_{i+2}-q_{i+1})\right]  -\psi_{i-1}^{+(1)}\exp\left[  2(q_{i}%
-q_{i-1})\right]  \right\}  ;
\end{align*}
This case admits a symmetric reduction, which reads%
\begin{align*}
\ddot{q}_{i}  &  =2\left\{  \left(  \psi_{i}^{(1)}\right)  ^{2}\exp\left[
2(q_{i+1}-q_{i})\right]  -\left(  \psi_{i-1}^{(1)}\right)  ^{2}\exp\left[
2(q_{i}-q_{i-1})\right]  \right\} \\
&  +2\left\{  \exp\left[  2(q_{i+2}-q_{i})\right]  -\exp\left[  2(q_{i}%
-q_{i-2})\right]  \right\}  ,\\
\dot{\psi}_{i}^{(1)}  &  =2\left(  \psi_{i+1}^{(1)}\exp\left[  2(q_{i+2}%
-q_{i+1})\right]  -\psi_{i-1}^{(1)}\exp\left[  2(q_{i}-q_{i-1})\right]
\right)  ;
\end{align*}

\item $m_{+}=3,$ $m_{-}=1$. We have%
\begin{align*}
\ddot{q}_{i}  &  =2\left\{  \psi_{i}^{+(1)}\exp\left[  2(q_{i+1}%
-q_{i})\right]  -\psi_{i-1}^{+(1)}\exp\left[  2(q_{i}-q_{i-1})\right]
\right\}  ,\\
\dot{\psi}_{i}^{+(1)}  &  =2\left\{  \psi_{i}^{+(2)}\exp\left[  2(q_{i+2}%
-q_{i+1})\right]  -\psi_{i-1}^{+(2)}\exp\left[  2(q_{i}-q_{i-1})\right]
\right\}  ,\\
\dot{\psi}_{i}^{+(2)}  &  =2\left\{  \exp\left[  2(q_{i+3}-q_{i+2})\right]
-\exp\left[  2(q_{i}-q_{i-1})\right]  \right\}  ;
\end{align*}

\item $m_{+}=3,$ $m_{-}=2$. The corresponding equations of motion read%
\begin{align*}
\ddot{q}_{i}  &  =2\left\{  \psi_{i}^{+(1)}\psi_{i}^{-(1)}\exp\left[
2(q_{i+1}-q_{i})\right]  -\psi_{i-1}^{-(1)}\psi_{i-1}^{+(1)}\exp\left[
2(q_{i}-q_{i-1})\right]  \right\} \\
&  +2\left\{  \psi_{i}^{+(2)}\psi_{i}^{-(2)}\exp\left[  2(q_{i+2}%
-q_{i})\right]  -\psi_{i-2}^{-(2)}\psi_{i-2}^{+(2)}\exp\left[  2(q_{i}%
-q_{i-2})\right]  \right\}  ,\\
\dot{\psi}_{i}^{+(1)}  &  =2\left\{  \psi_{i}^{+(2)}\psi_{i+1}^{-(1)}%
\exp\left[  2(q_{i+2}-q_{i+1})\right]  -\psi_{i-1}^{-(1)}\psi_{i-1}^{+(2)}%
\exp\left[  2(q_{i}-q_{i-1})\right]  \right\} \\
&  +2\left\{  \exp\left[  2(q_{i+3}-q_{i+1})\right]  -\exp\left[
2(q_{i}-q_{i-2})\right]  \right\}  ,\\
\dot{\psi}_{i}^{+(2)}  &  =2\left\{  \psi_{i+2}^{-(1)}\exp\left[
2(q_{i+3}-q_{i+2})\right]  -\psi_{i-1}^{-(1)}\exp\left[  2(q_{i}%
-q_{i-1})\right]  \right\}  ,\\
\dot{\psi}_{i}^{-(1)}  &  =2\left\{  \psi_{i+1}^{+(1)}\exp\left[
2(q_{i+2}-q_{i+1})\right]  -\psi_{i-1}^{+(1)}\exp\left[  2(q_{i}%
-q_{i-1})\right]  \right\}  ;
\end{align*}

\item $m_{+}=3,$ $m_{-}=3$. We have%
\begin{align*}
\ddot{q}_{i}  &  =2\left\{  \psi_{i}^{+(1)}\psi_{i}^{-(1)}\exp\left[
2(q_{i+1}-q_{i})\right]  -\psi_{i-1}^{+(1)}\psi_{i-1}^{-(1)}\exp\left[
2(q_{i}-q_{i-1})\right]  \right\} \\
&  +2\left\{  \psi_{i}^{+(2)}\psi_{i}^{-(2)}\exp\left[  2(q_{i+2}%
-q_{i})\right]  -\psi_{i-2}^{+(2)}\psi_{i-2}^{-(2)}\exp\left[  2(q_{i}%
-q_{i-2})\right]  \right\} \\
&  +2\left\{  \exp\left[  2(q_{i+3}-q_{i})\right]  -\exp\left[  2(q_{i}%
-q_{i-3})\right]  \right\}  ,\\
\dot{\psi}_{i}^{+(1)}  &  =2\left\{  \psi_{i}^{+(2)}\psi_{i+1}^{-(1)}%
\exp\left[  2(q_{i+2}-q_{i+1})\right]  -\psi_{i-1}^{-(1)}\psi_{i-1}^{+(2)}%
\exp\left[  2(q_{i}-q_{i-1})\right]  \right\} \\
&  +2\left\{  \psi_{i+1}^{-(2)}\exp\left[  2(q_{i+3}-q_{i+1})\right]
-\psi_{i-2}^{-(2)}\exp\left[  2(q_{i}-q_{i-2})\right]  \right\}  ,\\
\dot{\psi}_{i}^{+(2)}  &  =2\left\{  \psi_{i+2}^{-(1)}\exp\left[
2(q_{i+3}-q_{i+2})\right]  -\psi_{i-1}^{-(1)}\exp\left[  2(q_{i}%
-q_{i-1})\right]  \right\}  ,\\
\dot{\psi}_{i}^{-(1)}  &  =2\left\{  \psi_{i}^{-(2)}\psi_{i+1}^{+(1)}%
\exp\left[  2(q_{i+2}-q_{i+1})\right]  -\psi_{i-1}^{+(1)}\psi_{i-1}^{-(2)}%
\exp\left[  2(q_{i}-q_{i-1})\right]  \right\} \\
&  +2\left\{  \psi_{i+1}^{+(2)}\exp\left[  2(q_{i+3}-q_{i+1})\right]
-\psi_{i-2}^{+(2)}\exp\left[  2(q_{i}-q_{i-2})\right]  \right\}  ,\\
\dot{\psi}_{i}^{-(2)}  &  =2\left\{  \psi_{i+2}^{+(1)}\exp\left[
2(q_{i+3}-q_{i+2})\right]  -\psi_{i-1}^{+(1)}\exp\left[  2(q_{i}%
-q_{i-1})\right]  \right\}  ;
\end{align*}
This system of equations also admits a symmetric reduction, for which the
equations of motion read%
\begin{align*}
\ddot{q}_{i}  &  =2\left\{  \left(  \psi_{i}^{(1)}\right)  ^{2}\exp\left[
2(q_{i+1}-q_{i})\right]  -\left(  \psi_{i-1}^{(1)}\right)  ^{2}\exp\left[
2(q_{i}-q_{i-1})\right]  \right\} \\
&  +2\left\{  \left(  \psi_{i}^{(2)}\right)  ^{2}\exp\left[  2(q_{i+2}%
-q_{i})\right]  -\left(  \psi_{i-2}^{(2)}\right)  ^{2}\exp\left[
2(q_{i}-q_{i-2})\right]  \right\} \\
&  +2\left\{  \exp\left[  2(q_{i+3}-q_{i})\right]  -\exp\left[  2(q_{i}%
-q_{i-3})\right]  \right\}  ,\\
\dot{\psi}_{i}^{(1)}  &  =2\left(  \psi_{i}^{(2)}\psi_{i+1}^{(1)}\exp\left[
2(q_{i+2}-q_{i+1})\right]  -\psi_{i-1}^{(2)}\psi_{i-1}^{(1)}\exp\left[
2(q_{i}-q_{i-1})\right]  \right) \\
&  +2\left(  \psi_{i+1}^{(2)}\exp\left[  2(q_{i+3}-q_{i+1})\right]
-\psi_{i-2}^{(2)}\exp\left[  2(q_{i}-q_{i-2})\right]  \right)  ,\\
\dot{\psi}_{i}^{(2)}  &  =2\left(  \psi_{i+2}^{(1)}\exp\left[  2(q_{i+3}%
-q_{i+2})\right]  -\psi_{i-1}^{(1)}\exp\left[  2(q_{i}-q_{i-1})\right]
\right)  .
\end{align*}

\end{itemize}

\bigskip All these systems of equations are integrable in the Lax sense by
construction. However, to prove their Liouville integrability, we need to
truncate the number of degrees of freedom into a finite integer. There are two
possible ways to make the truncation, i.e. finite chain reduction and the
periodic reduction. We shall go into details of both kinds of reductions in
the following.

\subsection{Finite chain}

As just mentioned, the fact that equations (\ref{nonsym:1})-(\ref{nonsym:3})
contain an infinite number of variables makes it difficult to consider their
Liouville integrability. Therefore we now turn to a finite chain variant.

The Lax matrix $L$ for the finite chain case is given by ($m_{\pm}\leq n-1$)
\begin{align}
L  &  =L^{(0)}+L^{(>)}+L^{(<)},\nonumber\\
L^{(0)}  &  =%
%TCIMACRO{\dsum \limits_{i=1}^{n}}%
%BeginExpansion
{\displaystyle\sum\limits_{i=1}^{n}}
%EndExpansion
\dot{q}_{i}|i\rangle\langle i|,\nonumber\\
L^{(>)}  &  =%
%TCIMACRO{\dsum \limits_{k=1}^{m_{+}}}%
%BeginExpansion
{\displaystyle\sum\limits_{k=1}^{m_{+}}}
%EndExpansion%
%TCIMACRO{\dsum \limits_{i=1}^{n-k}}%
%BeginExpansion
{\displaystyle\sum\limits_{i=1}^{n-k}}
%EndExpansion
\exp(q_{i+k}-q_{i})\psi_{i}^{+(k)}|i\rangle\langle i+k|,\nonumber\\
L^{(<)}  &  =%
%TCIMACRO{\dsum \limits_{k=1}^{m_{-}}}%
%BeginExpansion
{\displaystyle\sum\limits_{k=1}^{m_{-}}}
%EndExpansion%
%TCIMACRO{\dsum \limits_{i=1}^{n-k}}%
%BeginExpansion
{\displaystyle\sum\limits_{i=1}^{n-k}}
%EndExpansion
\exp(q_{i+k}-q_{i})\psi_{i}^{-(k)}|i+k\rangle\langle i|. \label{Lfin}%
\end{align}
In other words, $L$ is obtained from $\mathcal{L}$ by setting $|i\rangle=0$
for $i<1$ and $i>n$. The corresponding $M$ matrix is again given by the
difference between $L^{(>)}$ and $L^{(<)}$,%
\[
M=L^{(>)}-L^{(<)}.
\]
The equations of motion now follows from a finite version of the Lax equation
(\ref{Lax}),
\begin{equation}
\dot{L}=[M,L], \label{Laxfin}%
\end{equation}
and it turns out that they again take the form of equations (\ref{nonsym:1}%
)-(\ref{nonsym:3}), but now the number of variables is restricted to be
finite, i.e.%
\begin{equation}%
\begin{array}
[c]{ccccc}%
q_{i}=0 & \mathrm{if} & i<1 & \mathrm{or} & i>n,\\
\psi_{i}^{\pm(k)}=0 & \mathrm{if} & i<1 & \mathrm{or} & i+k>n.
\end{array}
\label{fincond}%
\end{equation}
In what follows, we shall refer to the finite $(m_{+},m_{-})$-chain
characterized by the chain length $n$ as the $(m_{+},m_{-})$-th extended Toda
$n$-chain, or simply $(m_{+},m_{-})$-Toda $n$-chain. It should be remarked
that the finite (nonsymmetric) $(2,1)$- and $(2,2)$-chains respectively can be
regarded as dimensional reductions of the so-called heterotic Toda field
\cite{heterotic} and the bosonic superconformal Toda field \cite{bstf}
theories studied by one of the authors with others about ten years ago. On the
other hand, the particular cases of symmetrically reduced $(m,m)$-Toda
$n$-chain have already been discussed by Deift and collaborators \cite{Deift},
in particular, the symmetric $(2,2)$-chain was studied by them in detail.

The $(m_{+},m_{-})$-Toda $n$-chain is simpler than the $(m_{+},m_{-})$-th
extended infinite Toda chain (or the $(m_{+},m_{-})$-Toda $\infty$-chain) in
that its Liouville integrability can be established explicitly, and its
initial value problem can also be solved systematically. To show this we
remind that the operators $|i\rangle\langle j|$ may be identified with the
generator $e_{ij}$ of the Lie algebra $\mathfrak{gl}_{n}$. Therefore, we may
outline the Liouville integrability of the $(m_{+},m_{-})$-Toda $n$-chain
using the knowledge about the Lie algebra $\mathfrak{gl}_{n}$. The following
preposition is a straightforward generalization of the corresponding
preposition for the standard Toda chain (in the present terminology, these are
just the $(1,1)$-Toda $n$-chain), which is known to the integrable systems
community for over twenty years:

\begin{proposition}
Let $L$ be any matrix in $\mathfrak{gl}_{n}$ which has a strip form. Let $M$
be also a matrix in $\mathfrak{gl}_{n}$ which upper triangular part coincides
with that of $L$, while its lower triangular part differ from that of $L$ by a
total sign, i.e.
\[
M=L^{(>)}-L^{(<)}.
\]
Then, $L,$ $M$ satisfy the Lax equation (\ref{Laxfin}) and the quantities
$H_{k}(L):=\frac{1}{k}\mathrm{tr}(L^{k}),$ $k\in\mathbb{Z}$ are integrals of
motion for the matrix dynamics defined by the Lax equation.
\end{proposition}

For instance, the second integral of motion for the generic $(m_{+},m_{-}%
)$-Toda $n$-chain is easy to be calculated, yielding
\begin{align}
H_{2}(L)  &  =\frac{1}{2}\mathrm{tr}\left[  (L^{(0)})^{2}\right]
+\mathrm{tr}\left[  L^{(>)}L^{(<)}\right] \nonumber\\
&  =\frac{1}{2}\sum_{i=1}^{n}\dot{q}_{i}^{2}+\sum_{i=1}^{n-k}\sum_{k=1}%
^{m_{-}}\exp\left[  2\left(  q_{i+k}-q_{i}\right)  \right]  \psi_{i}%
^{+(k)}\psi_{i}^{-(k)}. \label{H2}%
\end{align}
The higher integrals of motion can also be constructed using the explicit
expression for $L$, but here we shall not list the explicit result since they
are quite complicated.

Moreover, by slightly generalizing the Hamiltonian description for the
standard Toda chain, we can see that the $L$ matrix satisfy the well known
fundamental Poisson relations for integrable systems, i.e.
\begin{equation}
\{L\otimes,L\}=[r^{\pm},L\otimes1+1\otimes L\}, \label{Poisson1}%
\end{equation}
where $r^{\pm}$ is the $r$ matrix for standard Toda chains, now written as%
\begin{align*}
r^{+}  &  =\sum_{i,j=1}^{n}|i\rangle\langle i|\otimes|j\rangle\langle
j|+\sum_{\substack{i,j=1\\i<j}}^{n}|i\rangle\langle j|\otimes|j\rangle\langle
i|,\\
r^{-}  &  =-\sum_{i,j=1}^{n}|i\rangle\langle i|\otimes|j\rangle\langle
j|-\sum_{\substack{i,j=1\\i>j}}^{n}|i\rangle\langle j|\otimes|j\rangle\langle
i|.
\end{align*}
This is achieved as follows. First, let us introduce the linear operator $R$
which acts on elements of $\mathfrak{gl}_{n}$ in the following way,%
\[
R(X)\equiv\mathrm{tr}_{2}(r^{+}+r^{-},1\otimes X),\quad\forall X\in
\mathfrak{gl}_{n},
\]
where $\mathrm{tr}_{2}$ means taking the trace over the second tensor
component of the tensor product space $\mathfrak{gl}_{n}\otimes\mathfrak{gl}%
_{n}$. Then for any linear function $f$ over $\mathfrak{gl}_{n}$, we can see
that the bracket \cite{Babelon}%
\begin{equation}
\{f(X),f(Y)\}\equiv f([X,Y]_{R}) \label{poissonR}%
\end{equation}
with%
\[
\lbrack X,Y]_{R}=[R(X),Y]+[X,R(Y)]
\]
is a Poisson bracket. In particular, if $f$ is chosen to be the Lax matrix
$L$, then the Poisson bracket (\ref{poissonR}) is exactly (\ref{Poisson1}). By
the way, we can see that $R(L)=L^{(>)}-L^{(<)}=M$. This establishes the
Hamiltonian description for the finite $(m_{+}, m_{-})$-Toda $n$-chain.

For the symmetrically reduced $n$-chain, it is easy to prove that the
following preposition holds, giving the unique solution to the initial value problem:

\begin{proposition}
The unique solution to the Lax system (\ref{Laxfin}) with symmetric matrix $L$
and initial data $L(0)=L_{0}$ is given by%
\[
L(t)=A(t)L_{0}A(t)^{-1},
\]
where $A(t)$ is the orthogonal component in the unique matrix decomposition%
\[
\exp(tL_{0})=A(t)N(t),
\]
where $N(t)$ is a lower triangular matrix.
\end{proposition}

For generic $(m_{+}, m_{-})$-Toda $n$-chain without symmetric reduction, the
solution to the initial value problem is quite complicated in practice.
However, the underlying idea is the same: one can make a decomposition for the
matrix $\exp(tL_{0})$ such that $\exp(tL_{0})=A(t)N(t)$ and $A(t)$ must
satisfy
\begin{equation}
\left(  \dot{A}(t)A^{-1}(t)\right)  ^{(>)} = \left(  A(t)L_{0}A^{-1}%
(t)\right)  ^{(>)}, \quad\left(  \dot{A}(t)A^{-1}(t)\right)  ^{(<)} = -\left(
A(t)L_{0}A^{-1}(t)\right)  ^{(<)}. \label{Acomp}%
\end{equation}
Then $L(t)=A(t)L_{0}A(t)^{-1}$ will solve the initial value problem with
initial data $L(0)=L_{0}$. The explicit solution for the matrix $A(t)$ can be
obtained by first writing $A(t)$ in the form of Gauss decomposition and then
solve (\ref{Acomp}) order by order in terms of the heights of the roots of
$\mathfrak{gl}_{n}$.

\subsection{Periodic chain}

The Lax matrix $L$ of the periodic $(m_{+},m_{-})$-Toda $n$-chain is given as
follows,%
\begin{align}
L  &  =L^{(0)}+L^{(>)}+L^{(<)},\nonumber\\
L^{(0)}  &  =%
%TCIMACRO{\dsum \limits_{i=0}^{n-1}}%
%BeginExpansion
{\displaystyle\sum\limits_{i=0}^{n-1}}
%EndExpansion
\dot{q}_{i}|i\rangle\langle i|,\nonumber\\
L^{(>)}  &  =%
%TCIMACRO{\dsum \limits_{k=1}^{m_{+}}}%
%BeginExpansion
{\displaystyle\sum\limits_{k=1}^{m_{+}}}
%EndExpansion%
%TCIMACRO{\dsum \limits_{i=0}^{n-k-1}}%
%BeginExpansion
{\displaystyle\sum\limits_{i=0}^{n-k-1}}
%EndExpansion
\exp(q_{i+k}-q_{i})\psi_{i}^{+(k)}|i\rangle\langle i+k|\nonumber\\
&  +w^{-1}%
%TCIMACRO{\dsum \limits_{k=1}^{m_{+}}}%
%BeginExpansion
{\displaystyle\sum\limits_{k=1}^{m_{+}}}
%EndExpansion%
%TCIMACRO{\dsum \limits_{i=n-k}^{n-1}}%
%BeginExpansion
{\displaystyle\sum\limits_{i=n-k}^{n-1}}
%EndExpansion
\exp(q_{i+k}-q_{i})\psi_{i}^{+(k)}|i\rangle\langle i-(n-k)|,\nonumber\\
L^{(<)}  &  =%
%TCIMACRO{\dsum \limits_{k=1}^{m_{-}}}%
%BeginExpansion
{\displaystyle\sum\limits_{k=1}^{m_{-}}}
%EndExpansion%
%TCIMACRO{\dsum \limits_{i=0}^{n-k-1}}%
%BeginExpansion
{\displaystyle\sum\limits_{i=0}^{n-k-1}}
%EndExpansion
\exp(q_{i+k}-q_{i})\psi_{i}^{-(k)}|i+k\rangle\langle i|\nonumber\\
&  +w%
%TCIMACRO{\dsum \limits_{k=1}^{m_{-}}}%
%BeginExpansion
{\displaystyle\sum\limits_{k=1}^{m_{-}}}
%EndExpansion%
%TCIMACRO{\dsum \limits_{i=n-k}^{n-1}}%
%BeginExpansion
{\displaystyle\sum\limits_{i=n-k}^{n-1}}
%EndExpansion
\exp(q_{i+k}-q_{i})\psi_{i}^{-(k)}|i-(n-k)\rangle\langle i|, \label{lperiod}%
\end{align}
where $w$ is a complex spectral parameter, and the $M$ matrix is the
difference between $L^{(>)}$ and $L^{(<)}$ given in (\ref{lperiod}),%
\[
M=L^{(>)}-L^{(<)}.
\]
The equations of motion following from the Lax equation still take the form of
(\ref{nonsym:1})-(\ref{nonsym:3}), but now the suffices of the variables
$q_{i}$ and $\psi_{i}^{\pm(k)}$ is to be understood modulo $n$, i.e.%
\begin{equation}
q_{i+rn}=q_{i},\quad\psi_{i+rn}^{\pm(k)}=\psi_{i}^{\pm(k)},\quad
i=0,1,...,n-1,\quad r\in\mathbb{Z}. \label{pericond}%
\end{equation}
Of course, just like the cases of infinite chain and finite $n$-chain, the
periodic $n$-chain also admits a symmetric reduction.

The Lax matrix (\ref{lperiod}) may be considered to be defined over the loop
algebra $\widetilde{\mathfrak{gl}}_{n}$. Its key difference from the case of
finite chains is the appearance of the complex spectral parameter $w$, which
can give rise to a complex curve called spectral curve. The Liouville
integrability and the Hamiltonian description for the periodic chains can be
outlined in a similar fashion like the case of finite chains, the only
difference being that we need to replace all the objects from the finite
dimensional Lie algebra $\mathfrak{gl}_{n}$ by objects from the loop algebra
$\widetilde{\mathfrak{gl}}_{n}$.

Among all periodic $(m_{+},m_{-})$ chains, the special cases of $(m_{+}%
,m_{-})=(1,1)$, $(2,1)$, $(2,2)$ are the simplest ones. The periodic
$(1,1)$-chain is just the standard Toda mechanics which we shall not pay any
more words. That the $(2,1)$- and $(2,2)$-chains are much simpler than the
generic cases is not only because that they contain less mechanical variables
but also due to the fact that the dynamics of these two special chains is
completely determined by the second integral of motion. More concretely we may
take the second integral of motion (notice that the lower and upper bounds for
the summations over $i$ are different from the finite chain cases)
\begin{align*}
H_{2}^{(2,1)}(L)  &  =\frac{1}{2}\sum_{i=0}^{n-1}\dot{q}_{i}^{2}+\sum
_{i=0}^{n-1}\exp\left[  2\left(  q_{i+1}-q_{i}\right)  \right]  \psi
_{i}^{+(1)},\quad\\
H_{2}^{(2,2)}(L)  &  =\frac{1}{2}\sum_{i=0}^{n-1}\dot{q}_{i}^{2}+\sum
_{i=0}^{n-1}\left\{  \exp\left[  2\left(  q_{i+1}-q_{i}\right)  \right]
\psi_{i}^{+(1)}\psi_{i}^{-(1)}+\exp\left[  2\left(  q_{i+2}-q_{i}\right)
\right]  \right\}
\end{align*}
as the Hamiltonian, and the equations of motion follow as Hamiltonian flows
for the variables $\dot{q}_{i},\psi_{i}^{\pm(1)}$ (recall that $\psi_{i}%
^{\pm(m_{\pm})}=1$) provided the nonzero Poisson brackets for these variables
are specified as%
\begin{align*}
\{q_{i},\dot{q}_{j}\}  &  =\delta_{ij},\\
\{\psi_{i}^{+(1)},\psi_{j}^{+(1)}\}  &  =2(\delta_{i,j-1}-\delta_{i,j+1}),\\
\{\psi_{i}^{-(1)},\psi_{j}^{-(1)}\}  &  =2(\delta_{i,j-1}-\delta_{i,j+1}).
\end{align*}
The Hamiltonian structure together with the Lax representation constitute a
description of the $(2,1)$- and $(2,2)$-chains as integrable Hamiltonian systems.

\section{Spectral problem of the periodic chain: an example}

As stated earlier in the introduction, a key object of interests in the study
of many body mechanical systems from the modern view point is the spectral
curves of the Lax matrix $L$, which we now study for the case of periodic
$(m_{+},m_{-})$-Toda $n$-chain.

First of all, the Lax equation can be viewed as the compatibility condition
for the pair of linear equations%
\begin{align}
\left(  \frac{d}{dt}-M\right)  \Phi &  =0,\label{pair1}\\
L\Phi &  =\lambda\Phi, \label{pair2}%
\end{align}
in which the second equation, (\ref{pair2}), defines the eigenvalue problem of
$L$.

On the other hand, since the mechanical variables appeared in $L$ defined in
(\ref{lperiod}) is periodic in the suffices with periodicity $n$, we can
define the monodromy operator $T$ which shifts the suffices of these
mechanical operators by $n$ and consequently acts on $q_{i}$ and $\psi
_{i}^{\pm(k)}$ as identity operator,%
\[
Tq_{i}=q_{i+n}=q_{i},\quad T\psi_{i}^{\pm(k)}=\psi_{i+n}^{\pm(k)}=\psi
_{i}^{\pm(k)}.
\]
However, the form of the $L$ matrix implies that $T$ acts on the vector states
$|i\rangle$ and $\langle i|$ respectively in a nontrivial way,%
\begin{align}
T|i\rangle &  =|i+n\rangle=w|i\rangle,\label{pered:1}\\
T\langle i|  &  =\langle i+n|=w^{-1}\langle i|, \label{pered:2}%
\end{align}
and hence the action of $T$ on $\Phi\equiv\sum_{i=0}^{n-1}\Phi_{i}|i\rangle$
is also nontrivial,%
\[
T\Psi=w\Psi.
\]
The actions of $L$ and $T$ on $\Psi$ commutes, $[L,T]=0$, which implies the
existence of certain relationship between $L$ and $T$,%
\[
\mathcal{P}(L,T)=0.
\]
Such a relationship can be formulated exactly by the spectral curve%
\[
\mathcal{P}(\lambda,w)=0,
\]
with the map $\mathcal{P}$ specified by the characteristic equation of the $L$
matrix,%
\begin{equation}
\mathcal{P}(\lambda,w)=\det\left(  L(w)-\lambda\right)  . \label{SpecMap}%
\end{equation}

The explicit form of the spectral map $\mathcal{P}(\lambda,w)$ for general
periodic $(m_{+},m_{-})$-Toda $n$-chain is very complicated, and we shall not
go into the detailed description for the generic cases. In stead, we shall
illustrate the structure of the spectral curve only for the simplest cases of
periodic $(2,1)$-chains.

For $(m_{+},m_{-})=(2,1)$, we can rewrite (\ref{SpecMap}) in explicit matrix
form,%
\begin{equation}
\mathcal{P}^{(2,1)}(\lambda,w)=\det\left(
\begin{array}
[c]{cccccc}%
\dot{q}_{0}-\lambda & a_{0}^{(1)} & a_{0}^{(2)} &  &  & wc_{n-1}^{(1)}\\
c_{0}^{(1)} & \dot{q}_{1}-\lambda & a_{1}^{(1)} & a_{1}^{(2)} &  & \\
& c_{1}^{(1)} & \dot{q}_{2}-\lambda & \ddots & \ddots & \\
&  & \ddots & \ddots & \ddots & a_{n-3}^{(2)}\\
w^{-1}a_{n-2}^{(2)} &  &  & \ddots & \ddots & a_{n-2}^{(1)}\\
w^{-1}a_{n-1}^{(1)} & w^{-1}a_{n-1}^{(2)} &  &  & c_{n-2}^{(1)} & \dot
{q}_{n-1}-\lambda
\end{array}
\right)  , \label{spectra21}%
\end{equation}
with entries $a_{i}^{(k)}$, $c_{i}^{(k)}$ given by equation (\ref{subs}),
where suffices for the variables $q_{i}$, $\psi_{i}^{\pm(k)}$ are to be
understood modulo $n$. A direct evaluation of (\ref{spectra21}) yields%
\begin{equation}
\mathcal{P}^{(2,1)}(\lambda,w)=(-1)^{n-1}w+\frac{Q_{n-3}(\lambda)}{w}+\frac
{1}{w^{2}}+P_{n}(\lambda)=0, \label{21spect}%
\end{equation}
where $Q_{n-3}(\lambda)$ and $P_{n}(\lambda)$ respectively are some
polynomials in $\lambda$ of order $n-3$ and $n$, with coefficients being
polynomials in $\dot{q}_{i},\exp(q_{i})$ and $\psi_{i}^{+(1)}$ which are
symmetric under exchange of suffices. It should be remarked that the form of
the spectral curve (\ref{21spect}) for $(2,1)$-Toda chains is very different
fron that of the $(1,1)$-chains. Therefore, it is interesting to learn what
kind of SUSY gauge theories might correspond, taking (\ref{21spect}) as its
Seiberg-Witten curve. We leave this problem for future study.

Another way of looking at the spectral problems for the $(m_{+},m_{-})$-Toda
chains can be specified as follows. First, let us define an analogous
eigenvalue problem to eq. (\ref{pair2}) before making the periodic reduction.
This amounts to introduce formally an eigenvector $\Phi=\sum_{i\in\mathbb{Z}%
}\Phi_{i}|i\rangle$ for the $\mathcal{L}$ matrix (\ref{L}), i.e.
\begin{equation}
\mathcal{L}\Phi=\lambda\Phi. \label{eigeninf}%
\end{equation}
Using the explicit form of $\mathcal{L}$ we can rewrite (\ref{eigeninf}) as%
\begin{align}%
%TCIMACRO{\dsum \limits_{i\in\mathbb{Z}}}%
%BeginExpansion
{\displaystyle\sum\limits_{i\in\mathbb{Z}}}
%EndExpansion
\lambda\Phi_{i}|i\rangle &  =%
%TCIMACRO{\dsum \limits_{i\in\mathbb{Z}}}%
%BeginExpansion
{\displaystyle\sum\limits_{i\in\mathbb{Z}}}
%EndExpansion
\dot{q}_{i}\Phi_{i}|i\rangle\nonumber\\
&  +%
%TCIMACRO{\dsum \limits_{k=1}^{m_{+}}}%
%BeginExpansion
{\displaystyle\sum\limits_{k=1}^{m_{+}}}
%EndExpansion%
%TCIMACRO{\dsum \limits_{i\in\mathbb{Z}}}%
%BeginExpansion
{\displaystyle\sum\limits_{i\in\mathbb{Z}}}
%EndExpansion
\exp(q_{i+k}-q_{i})\psi_{i}^{+(k)}\Phi_{i+k}|i\rangle\nonumber\\
&  +%
%TCIMACRO{\dsum \limits_{k=1}^{m_{-}}}%
%BeginExpansion
{\displaystyle\sum\limits_{k=1}^{m_{-}}}
%EndExpansion%
%TCIMACRO{\dsum \limits_{i\in\mathbb{Z}}}%
%BeginExpansion
{\displaystyle\sum\limits_{i\in\mathbb{Z}}}
%EndExpansion
\exp(q_{i}-q_{i-k})\psi_{i-k}^{-(k)}\Phi_{i-k}|i\rangle. \label{GenRecur}%
\end{align}
Since the states $|i\rangle$ are all linearly independent, the last equation
is in fact a collection of equations which provide a recursive relationship
between the vector components $\Phi_{i}$, i.e.
\begin{align}
\Phi_{i+m_{+}}  &  =\left(  \lambda-\dot{q}_{i}\right)  \Phi_{i}-%
%TCIMACRO{\dsum \limits_{k=1}^{m_{+}-1}}%
%BeginExpansion
{\displaystyle\sum\limits_{k=1}^{m_{+}-1}}
%EndExpansion
\exp(q_{i+k}-q_{i+m_{+}})\psi_{i}^{+(k)}\Phi_{i+k}\nonumber\\
&  \quad-%
%TCIMACRO{\dsum \limits_{k=1}^{m_{-}}}%
%BeginExpansion
{\displaystyle\sum\limits_{k=1}^{m_{-}}}
%EndExpansion
\exp(2q_{i}-q_{i-k}-q_{i+m_{+}})\psi_{i-k}^{-(k)}\Phi_{i-k},\quad
i\in\mathbb{Z}. \label{recureq}%
\end{align}
Therefore, for generic values of $m_{+}$ and $m_{-}$, such a set of equations
can be recast into matrix form with coefficient matrix of dimension
$m_{+}+m_{-}$. More concretely, writing%
\[
\Theta_{i}=\left(
\begin{array}
[c]{ccccc}%
\Phi_{i+m_{+}-1} & \Phi_{i+m_{+}-2} & \cdots & \Phi_{i-m_{-}+1} &
\Phi_{i-m_{-}}%
\end{array}
\right)  ^{T}%
\]
and defining the recursive matrix $\mathcal{L}_{i}$ in terms of the matrix
equation%
\[
\Theta_{i+1}=\mathcal{L}_{i}\Theta_{i},
\]
we may read out the matrix $\mathcal{L}_{i}$ from (\ref{recureq}),%
\[
\mathcal{L}_{i}=\left(
\begin{array}
[c]{ccccc}%
\ast & \ast & \ast & \cdots & \ast\\
1 & 0 &  & \cdots & 0\\
0 & 1 & 0 &  & \\
\vdots & \ddots & \ddots & \ddots & \vdots\\
0 & \cdots & 0 & 1 & 0
\end{array}
\right)  ,
\]
where the $\ast$'s are just the coefficients of $\Phi_{j}$'s appeared in
(\ref{recureq}). The form of the above matrices $\mathcal{L}_{i}$ is very
similar to the Lax matrix of the KdV hierarchy in the Drinfeld-Sokolov gauge
\cite{DS}, which reminds us that it may be gauge transformed into a form with
the only nontrivial elements (i.e. the ones besides the 1's on the
subdiagonal) being on the main diagonal following the standard procedure of
Miura transformation. However, at present, we shall not need to make such a transformation.

Now acting on, say, $\Theta_{0}$, from the left by $\mathcal{L}_{0}$ and then
by $\mathcal{L}_{1}$, $\mathcal{L}_{2}$, ... until $\mathcal{L}_{n-1}$, we
have%
\[
\Theta_{n}=\mathcal{L}_{n-1}\mathcal{L}_{n-2}\cdots\mathcal{L}_{0}\Theta_{0}.
\]
It is now a good place to recall the periodic reduction condition
(\ref{pered:1}) which implies the relationship%
\[
\Theta_{n}=w\Theta_{0}.
\]
Writing
\begin{equation}
T(\lambda)=\mathcal{L}_{n-1}\mathcal{L}_{n-2}\cdots\mathcal{L}_{0}(\lambda)
\label{T}%
\end{equation}
where explicit dependence on the parameter $\lambda$ is emplasized, we have%
\[
T(\lambda)\Theta_{0}=w\Theta_{0}.
\]
We thus have an alternative spectral problem%
\begin{equation}
\det\left(  T(\lambda)-w\right)  =0. \label{spectraalt}%
\end{equation}
This last equation gives another expression for the spectral curves associated
with the periodic $(m_{+},m_{-})$-Toda $n$-chain.

As an illustrating example, we shall give the explicit form of the matrices
$\mathcal{L}_{i}$ for the case of $(m_{+},m_{-})=(2,1)$. In this case, we have%
\[
\Theta_{i}^{(2,1)}=\left(
\begin{array}
[c]{c}%
\Phi_{i+1}\\
\Phi_{i}\\
\Phi_{i-1}%
\end{array}
\right)  ,\quad
\]
and consequently
\[
\mathcal{L}_{i}^{(2,1)}=\left(
\begin{array}
[c]{ccc}%
-\exp(q_{i+1}-q_{i+2})\psi_{i}^{+(1)} & \left(  \lambda-\dot{q}_{i}\right)
\exp(q_{i}-q_{i+2}) & -\exp(2q_{i}-q_{i-1}-q_{i+2})\\
1 & 0 & 0\\
0 & 1 & 0
\end{array}
\right)  .
\]
It is then a simple practice to get $T^{(2,1)}(\lambda)$ using (\ref{T}), and
then get the spectral curve via (\ref{spectraalt}). That the two spectral
curves (\ref{SpecMap}) and (\ref{spectraalt}) are identical is a well known
fact for the case of standard Toda chains (i.e. the $(1,1)$-chains).

\section{Nonabelian generalizations}

The extended Toda chain we studied so far can also be generalized along
another line, i.e. the nonabelian generalizations. By nonabelian
generalization we mean that the fundamental mechanical variables $q_{i}$,
$a_{i}^{(k)}$ and $c_{i}^{(k)}$ can be replaced by nonabelian objects, e.g.
$q_{i}$ may be replaced by $\ell_{i}\times$ $\ell_{i}$ matrices $Q_{i}$, while
$a_{i}^{(k)}$ and $c_{i}^{(k)}$ are to be replaced by some generic rectangular
matrices $A_{i}^{(k)}$ of size $\ell_{i}\times$ $\ell_{i+k}$ and $C_{i}^{(k)}$
of size $\ell_{i+k}\times$ $\ell_{i}$. Such nonabelian versions of the
extended Toda chain is expected to be useful in the the study of moduli
structure of $N=2$ and $N=1$ SUSY gauge theories with non-semisimple or
partially broken gauge symmetries.

Now let us give explicitly the Lax matrices and the equations of motion for
the nonabelian extended Toda chain. For the case of infinite chain, the
$\mathcal{L}$ matrix is given as%
\begin{align*}
\mathcal{L}  &  =\mathcal{L}^{(0)}+\mathcal{L}^{(>)}+\mathcal{L}^{(<)},\\
\mathcal{L}^{(0)}  &  =%
%TCIMACRO{\dsum \limits_{i\in\mathbb{Z}}}%
%BeginExpansion
{\displaystyle\sum\limits_{i\in\mathbb{Z}}}
%EndExpansion
\dot{Q}_{i}|i\rangle\langle i|,\\
\mathcal{L}^{(>)}  &  =%
%TCIMACRO{\dsum \limits_{i\in\mathbb{Z}}}%
%BeginExpansion
{\displaystyle\sum\limits_{i\in\mathbb{Z}}}
%EndExpansion%
%TCIMACRO{\dsum \limits_{k=1}^{m_{+}}}%
%BeginExpansion
{\displaystyle\sum\limits_{k=1}^{m_{+}}}
%EndExpansion
A_{i}^{(k)}|i\rangle\langle i+k|,\\
\mathcal{L}^{(<)}  &  =%
%TCIMACRO{\dsum \limits_{i\in\mathbb{Z}}}%
%BeginExpansion
{\displaystyle\sum\limits_{i\in\mathbb{Z}}}
%EndExpansion%
%TCIMACRO{\dsum \limits_{k=1}^{m_{-}}}%
%BeginExpansion
{\displaystyle\sum\limits_{k=1}^{m_{-}}}
%EndExpansion
C_{i}^{(k)}|i+k\rangle\langle i|.
\end{align*}
The $\mathcal{M}$ matrix is still the difference between $\mathcal{L}^{(>)}$
and $\mathcal{L}^{(<)}$,%
\[
\mathcal{M=L}^{(>)}-\mathcal{L}^{(<)}.
\]
The equations of motion follow directly from the Lax equation $\mathcal{\dot
{L}=[M},\mathcal{L]}$ and they read%
\begin{align}
\ddot{Q}_{i}  &  =2%
%TCIMACRO{\dsum \limits_{k=1}^{m_{-}}}%
%BeginExpansion
{\displaystyle\sum\limits_{k=1}^{m_{-}}}
%EndExpansion
\left\{  A_{i}^{(k)}C_{i}^{(k)}-C_{i-k}^{(k)}A_{i-k}^{(k)}\right\}
,\label{nonabeq1}\\
\dot{A}_{i}^{(k)}  &  =A_{i}^{(k)}\dot{Q}_{i+k}-\dot{Q}_{i}A_{i}%
^{(k)}\nonumber\\
&  +2%
%TCIMACRO{\dsum \limits_{k^{\prime}=1}^{\min(m_{+}-k,m_{-})}}%
%BeginExpansion
{\displaystyle\sum\limits_{k^{\prime}=1}^{\min(m_{+}-k,m_{-})}}
%EndExpansion
\left(  A_{i}^{(k+k^{\prime})}C_{i+k}^{(k^{\prime})}-C_{i-k^{\prime}%
}^{(k^{\prime})}A_{i-k^{\prime}}^{(k+k^{\prime})}\right)  ,\qquad(1\leq k\leq
m_{+}-1)\label{nonabeq2}\\
\dot{C}_{i}^{(k)}  &  =\dot{Q}_{i+k}C_{i}^{(k)}-C_{i}^{(k)}\dot{Q}%
_{i}\nonumber\\
&  +2%
%TCIMACRO{\dsum \limits_{k^{\prime}=1}^{m_{-}-k}}%
%BeginExpansion
{\displaystyle\sum\limits_{k^{\prime}=1}^{m_{-}-k}}
%EndExpansion
\left(  A_{i+k}^{(k^{\prime})}C_{i}^{(k+k^{\prime})}-C_{i-k^{\prime}%
}^{(k+k^{\prime})}A_{i-k^{\prime}}^{(k^{\prime})}\right)  ,\qquad(1\leq k\leq
m_{-}-1)\label{nonabeq3}\\
\dot{A}_{i}^{(m_{+})}  &  =A_{i}^{(m_{+})}\dot{Q}_{i+m_{+}}-\dot{Q}_{i}%
A_{i}^{(m_{+})},\label{nonabeq4}\\
\dot{C}_{i}^{(m_{-})}  &  =\dot{Q}_{i+m_{-}}C_{i}^{(m_{-})}-C_{i}^{(m_{-}%
)}\dot{Q}_{i}. \label{nonabeq5}%
\end{align}
The Lax integrability for the system of equations (\ref{nonabeq1}%
)-(\ref{nonabeq5}) is guaranteed by construction. However, for generic matrix
variables $Q_{i},A_{i}^{(k)}$ and $C_{i}^{(k)}$, the form of these equations
cannot be simplified any further. In particular, the appearance of $\dot
{Q}_{i}$ on the right hand side of eqs. (\ref{nonabeq2})-(\ref{nonabeq5}) is inevitable.

Although the construction given above corresponds to infinite nonabelian
chains only, it is extremely easy to get the equations of motion for the
finite and periodic chains out of (\ref{nonabeq1})-(\ref{nonabeq5}) by
imposing constraints over the mechanical variables $Q_{i},$ $A_{i}^{(k)}$ and
$C_{i}^{(k)}$ following the similar fashion as made in eqs. (\ref{fincond})
and (\ref{pericond}). Concretely, the equations of motion for the finite
nonabelian $(m_{+},m_{-})$-Toda $n$-chain are given by (\ref{nonabeq1}%
)-(\ref{nonabeq5}) together with the constraint conditions%
\[%
\begin{array}
[c]{ccccc}%
Q_{i}=0 & \mathrm{if} & i<1 & \mathrm{or} & i>n,\\
A_{i}^{(k)}=C_{i}^{(k)}=0 & \mathrm{if} & i<1 & \mathrm{or} & i+k>n.
\end{array}
\]
The equations of motion for the periodic $(m_{+},m_{-})$-Toda $n$-chain are
also given by (\ref{nonabeq1})-(\ref{nonabeq5}) together with the following
constraint conditions,%
\[
Q_{i+rn}=Q_{i},\quad A_{i+rn}^{(k)}=A_{i}^{(k)},\quad C_{i+rn}^{(k)}%
=C_{i}^{(k)},\quad r\in\mathbb{Z}.
\]
For the size of the matrices to be consistent, we must also require
\[
\ell_{i+rn}=\ell_{i},\quad r\in\mathbb{Z}.
\]

In order that the nonabelian generalizations take a similar form as
(\ref{nonsym:1})-(\ref{nonsym:3}), we now consider to replace the matrices
$\dot{Q}_{i}$ by certain symmetric right invariant vector field over
$GL(\ell_{i})$, namely%
\[
\dot{Q}_{i}\rightarrow\dot{B}_{i}B_{i}^{-1},\qquad\left(  \dot{B}_{i}%
B_{i}^{-1}\right)  ^{T}=\dot{B}_{i}B_{i}^{-1},\qquad B_{i}\in GL(\ell_{i}).
\]
Such a replacement of variables is a kind of reduction to the original system
of equations, and hence the resulting system%
\begin{align*}
\frac{d}{dt}\left(  \dot{B}_{i}B_{i}^{-1}\right)   &  =2%
%TCIMACRO{\dsum \limits_{k=1}^{m_{-}}}%
%BeginExpansion
{\displaystyle\sum\limits_{k=1}^{m_{-}}}
%EndExpansion
\left\{  A_{i}^{(k)}C_{i}^{(k)}-C_{i-k}^{(k)}A_{i-k}^{(k)}\right\}  ,\\
\frac{d}{dt}\left(  A_{i}^{(k)}\right)  ^{T}  &  =\left(  \dot{B}_{i+k}%
B_{i+k}^{-1}\right)  \left(  A_{i}^{(k)}\right)  ^{T}-\left(  A_{i}%
^{(k)}\right)  ^{T}\left(  \dot{B}_{i}B_{i}^{-1}\right) \\
&  +2%
%TCIMACRO{\dsum \limits_{k^{\prime}=1}^{\min(m_{+}-k,m_{-})}}%
%BeginExpansion
{\displaystyle\sum\limits_{k^{\prime}=1}^{\min(m_{+}-k,m_{-})}}
%EndExpansion
\left\{  \left(  C_{i+k}^{(k^{\prime})}\right)  ^{T}\left(  A_{i}%
^{(k+k^{\prime})}\right)  ^{T}-\left(  A_{i-k^{\prime}}^{(k+k^{\prime}%
)}\right)  ^{T}\left(  C_{i-k^{\prime}}^{(k^{\prime})}\right)  ^{T}\right\}
,\\
&  (1\leq k\leq m_{+}-1)\\
\frac{d}{dt}C_{i}^{(k)}  &  =\left(  \dot{B}_{i+k}B_{i+k}^{-1}\right)
C_{i}^{(k)}-C_{i}^{(k)}\left(  \dot{B}_{i}B_{i}^{-1}\right) \\
&  +2%
%TCIMACRO{\dsum \limits_{k^{\prime}=1}^{m_{-}-k}}%
%BeginExpansion
{\displaystyle\sum\limits_{k^{\prime}=1}^{m_{-}-k}}
%EndExpansion
\left(  A_{i+k}^{(k^{\prime})}C_{i}^{(k+k^{\prime})}-C_{i-k^{\prime}%
}^{(k+k^{\prime})}A_{i-k^{\prime}}^{(k^{\prime})}\right)  ,\qquad(1\leq k\leq
m_{-}-1)\\
\frac{d}{dt}\left(  A_{i}^{(m_{+})}\right)  ^{T}  &  =\left(  \dot{B}%
_{i+m_{+}}B_{i+m_{+}}^{-1}\right)  \left(  A_{i}^{(m_{+})}\right)
^{T}-\left(  A_{i}^{(m_{+})}\right)  ^{T}\left(  \dot{B}_{i}B_{i}^{-1}\right)
,\\
\frac{d}{dt}C_{i}^{(m_{-})}  &  =\left(  \dot{B}_{i+m_{-}}B_{i+m_{-}}%
^{-1}\right)  C_{i}^{(m_{-})}-C_{i}^{(m_{-})}\left(  \dot{B}_{i}B_{i}%
^{-1}\right)
\end{align*}
is not strictly equivalent to the original one. In the last equations, we used
the symbol $X^{T}$ to denote the transpose of the matrix $X$. After the above
reduction, we may further let%
\[
\left(  A_{i}^{(k)}\right)  ^{T}=B_{i+k}\left(  \Psi_{i}^{+(k)}\right)
^{T}B_{i}^{-1},\qquad C_{i}^{(k)}=B_{i+k}\Psi_{i}^{-(k)}B_{i}^{-1},
\]
so that in terms of the new variables $B_{i}$, $\Psi_{i}^{+(k)}$ and $\Psi
_{i}^{-(k)}$, the reduced system of equations reads%
\begin{align}
\frac{d}{dt}\left(  \dot{B}_{i}B_{i}^{-1}\right)   &  =2%
%TCIMACRO{\dsum \limits_{k=1}^{m_{-}}}%
%BeginExpansion
{\displaystyle\sum\limits_{k=1}^{m_{-}}}
%EndExpansion
\left\{  \left(  B_{i}^{-1}\right)  ^{T}\Psi_{i}^{+(k)}\left(  B_{i+k}\right)
^{T}B_{i+k}\Psi_{i}^{-(k)}B_{i}^{-1}\right. \nonumber\\
&  \left.  -B_{i}\Psi_{i-k}^{-(k)}B_{i-k}^{-1}\left(  B_{i-k}^{-1}\right)
^{T}\Psi_{i-k}^{+(k)}\left(  B_{i}\right)  ^{T}\right\}  ,\\
\frac{d}{dt}\Psi_{i}^{+(k)}  &  =2\left(  B_{i}\right)  ^{T}%
%TCIMACRO{\dsum \limits_{k^{\prime}=1}^{\min(m_{+}-k,m_{-})}}%
%BeginExpansion
{\displaystyle\sum\limits_{k^{\prime}=1}^{\min(m_{+}-k,m_{-})}}
%EndExpansion
\left\{  \left(  B_{i}^{-1}\right)  ^{T}\Psi_{i}^{+(k+k^{\prime})}\left(
B_{i+k+k^{\prime}}\right)  ^{T}B_{i+k+k^{\prime}}\Psi_{i+k}^{-(k^{\prime}%
)}B_{i+k}^{-1}\right. \nonumber\\
\quad &  \left.  -B_{i}\Psi_{i-k^{\prime}}^{-(k^{\prime})}B_{i-k^{\prime}%
}^{-1}\left(  B_{i-k^{\prime}}^{-1}\right)  ^{T}\Psi_{i-k^{\prime}%
}^{+(k+k^{\prime})}\left(  B_{i+k}\right)  ^{T}\right\}  \left(  \left(
B_{i+k}\right)  ^{-1}\right)  ^{T},\qquad\nonumber\\
&  (1\leq k\leq m_{+}-1)\\
\frac{d}{dt}\Psi_{i}^{-(k)}  &  =2\left(  B_{i+k}\right)  ^{-1}%
%TCIMACRO{\dsum \limits_{k^{\prime}=1}^{m_{-}-k}}%
%BeginExpansion
{\displaystyle\sum\limits_{k^{\prime}=1}^{m_{-}-k}}
%EndExpansion
\left\{  \left(  B_{i+k}^{-1}\right)  ^{T}\Psi_{i+k}^{+(k^{\prime})}\left(
B_{i+k+k^{\prime}}\right)  ^{T}B_{i+k+k^{\prime}}\Psi_{i}^{-(k+k^{\prime}%
)}B_{i}^{-1}\right. \nonumber\\
&  \left.  -B_{i+k}\Psi_{i-k^{\prime}}^{-(k+k^{\prime})}B_{i-k^{\prime}}%
^{-1}\left(  B_{i-k^{\prime}}^{-1}\right)  ^{T}\Psi_{i-k^{\prime}%
}^{+(k^{\prime})}\left(  B_{i}\right)  ^{T}\right\}  B_{i},\qquad\nonumber\\
&  (1\leq k\leq m_{-}-1)\\
\frac{d}{dt}\Psi_{i}^{+(m_{+})}  &  =0,\qquad\frac{d}{dt}\Psi_{i}^{-(m_{-}%
)}=0.
\end{align}

We may also consider the nonabelian analogues of the symmetric reductions when
$m_{+}=m_{-}=m$. The condition for the symmetric reduction reads%
\[
\Psi_{i}^{+(k)}=\left(  \Psi_{i}^{-(k)}\right)  ^{T}\equiv\Psi_{i}^{(k)},
\]
and the reduced equations read%
\begin{align}
\frac{d}{dt}\left(  \dot{B}_{i}B_{i}^{-1}\right)   &  =2%
%TCIMACRO{\dsum \limits_{k=1}^{m}}%
%BeginExpansion
{\displaystyle\sum\limits_{k=1}^{m}}
%EndExpansion
\left\{  \left(  B_{i}^{-1}\right)  ^{T}\Psi_{i}^{(k)}\left(  B_{i+k}\right)
^{T}B_{i+k}\left(  \Psi_{i}^{(k)}\right)  ^{T}B_{i}^{-1}\right. \nonumber\\
&  \left.  -B_{i}\left(  \Psi_{i-k}^{(k)}\right)  ^{T}B_{i-k}^{-1}\left(
B_{i-k}^{-1}\right)  ^{T}\Psi_{i-k}^{(k)}\left(  B_{i}\right)  ^{T}\right\}
,\\
\frac{d}{dt}\Psi_{i}^{(k)}  &  =2\left(  B_{i}\right)  ^{T}%
%TCIMACRO{\dsum \limits_{k^{\prime}=1}^{m-k}}%
%BeginExpansion
{\displaystyle\sum\limits_{k^{\prime}=1}^{m-k}}
%EndExpansion
\left\{  \left(  B_{i}^{-1}\right)  ^{T}\Psi_{i}^{(k+k^{\prime})}\left(
B_{i+k+k^{\prime}}\right)  ^{T}B_{i+k+k^{\prime}}\left(  \Psi_{i+k}%
^{(k^{\prime})}\right)  ^{T}B_{i+k}^{-1}\right. \nonumber\\
\quad &  \left.  -B_{i}\left(  \Psi_{i-k^{\prime}}^{(k^{\prime})}\right)
^{T}B_{i-k^{\prime}}^{-1}\left(  B_{i-k^{\prime}}^{-1}\right)  ^{T}%
\Psi_{i-k^{\prime}}^{(k+k^{\prime})}\left(  B_{i+k}\right)  ^{T}\right\}
\left(  \left(  B_{i+k}\right)  ^{-1}\right)  ^{T},\qquad\nonumber\\
&  (1\leq k\leq m-1)\\
\frac{d}{dt}\Psi_{i}^{(m)}  &  =0.
\end{align}

It is also interesting to study the spectral curves and stationary point
configurations for the nonabelian generalizations. However, the corresponding
calculations would be much more complicated than the abelian cases, and we
shall not go into any details along this direction.

\section{Conclusion and discussion}

In this paper, we studied the integrability of generalized Toda chains based
on the Lie algebras $\mathfrak{gl}_{n}$ (including the case $n=\infty$) and
$\widetilde{\mathfrak{gl}}_{n}$ in full generality. For each one of these Lie
algebras we find a family of integrable generalizations of Toda chains
characterized by two integers $m_{+}$ and $m_{-}$, with the Lax matrices and
equations of motion given explicitly. For the case of $m_{+}=m_{-}$ there is a
symmetric reduction for the equations of motion, and the solution to the
initial value problem is outlined. For each pair of $m_{+}$ and $m_{-},$we
also constructed a family of nonabelian generalizations of the Toda chains.

Another problem that we considered in this paper, but not in full generality,
is the spectral curves of the abelian periodic $(m_{+},m_{-})$-Toda chains. We
outlined two different ways for the description of the spectral problems, and
presented the form of the spectral curve in the special case of $(2,1)$-chains.

We should remark that for $n=\infty$, the (abelian) generalized Toda chains we
get may be considered as dimensional reductions of the well-known Toda lattice
hierarchy \cite{Ueno}, which is a collection of infinite many integrable field
theoretic models in (1+1)-dimensions. However, the equations of motion for the
Toda lattice hierarchy were never written in such an explicit and unified way
as we did in this paper.

Although the general construction we made in this paper is based on the Lie
algebras $\mathfrak{gl}_{n}$ and $\widetilde{\mathfrak{gl}}_{n}$ only, the
results is ready to be applied to the simple Lie algebras $\mathfrak{sl}_{n}$
and $\widetilde{\mathfrak{sl}}_{n}$ by simply imposing a traceless condition
on the $L$ matrices. Further generalizations to the cases of $\mathfrak{so}%
_{n}$, $\mathfrak{sp}_{n}$ and their loop algebras is also possible, which we
leave for later study. Another important problem which we also leave for
future study is the application of spectral curves for the periodic
$(m_{+},m_{-})$-Toda chains in $n=2$ SUSY gauge theories. The stationary point
configurations for the periodic chains is also of great interests from the
$N=1$ gauge theoretic point of view.

\section*{Acknowledgement}

LZ is grateful to Ian MacIntosh for introducing to him the technique for the
construction of initial value problem for the standard periodic Toda chain and
bring his attention to \cite{Deift}. This work in supported in part by the NSFC.


\begin{thebibliography}{99}                                                                                               %


\bibitem {1}M. Toda, ``Vibration of a chain with nonlinear interaction", J.
Phys. Soc. Japan, Vol.22, (1967) 431-436; ``Wave Propagation in Anharmonic
Lattices", J. Phys. Soc. Japan, Vol.23, (1967) 501-506.

\bibitem {2}H. Flaschka, ``The Toda Lattice II. Existence of integrals" Phys.
Rev.B9 (1974) 1924-1925; ``On the Toda Lattice II. Inverse Scattering
Solution", Prog. Theor. Phys.51 (1974) 703-716.

\bibitem {3}F. Calogero, ``Solution of the one-dimensional N-body problems
with quadratic and/or inversely quadratic pair potentials", J. Math. Phys. 12,
419-436 (1971); Erratum, ibidem 37 (1996) 3646 ; ``Exactly solvable
one-dimensional many-body problems", Lett. Nuovo Cimento 13, 411-416 (1975).

\bibitem {4}J. Moser, ``Three integrable Hamiltonian systems connected with
isospectral deformations", Adv. Math. 16 (1975) 197-220 ; ``Various aspects of
integrable Hamiltonian systems", in: Dynamical Systems, Prog. in Math.8,
Birkhauser, Basel, 1980, pp. 233-289.

\bibitem {8}S. N. M. Ruijsenaars and H. Schneider, ``A new class of integrable
systems and its relation to solitons", Ann. Phys. (NY) 170 (1986) 370-405.

\bibitem {7}S. N. M. Ruijsenaars, ``Complete integrability of relativistic
Calogero- Moser systems and elliptic function identities", Commun. Math. Phys.
110 (1987) 191-213.

\bibitem {5}R.Donagi and E.Witten, ``Supersymmetric Yang-Mills Systems And
Integrable Systems'', Nucl. Phys. B460 (1996) 299-334, hep-th/9510101.

\bibitem {6}See, e.g. A. Marshakov, ``Seiberg-Witten Theory and Integrable
Systems'', World Scientific Publishing, 1999 and references therein.

\bibitem {9}N. Seiberg and E. Witten, ``Monopole Condensation, And Confinement
In N=2 Supersymmetric Yang-Mills Theory'', Nucl. Phys.B426 (1994) 19-52, hep-th/9407087.

\bibitem {DV}R. Dijkgraaf and C. Vafa, ``Matrix Models, Topological Strings,
and Supersymmetric Gauge Theories'', Nucl.Phys. B644 (2002) 3-20,
hep-th/0206255; ``On Geometry and Matrix Models'', Nucl.Phys. B644 (2002)
21-39, hep-th/0207106; ``A Perturbative Window into Non-Perturbative
Physics'', hep-th/0208048.

\bibitem {Dorey}N.Dorey, ``An Elliptic Superpotential for Softly Broken N=4
Supersymmetric Yang-Mills Theory'', JHEP 9907 (1999) 021, hep-th/9906011.

\bibitem {heterotic}L. Chao and B-Y. Hou, ``Heterotic Toda fields", Nucl.
Phys. B436 (1995) 638-658.

\bibitem {bstf}B-Y. Hou and L. Chao, ``From integrability to conformality:
Bosonic superconformal Toda theories", Int. J. Mod. Phys. A8(6) (1993) 1105-1123.

\bibitem {Deift}P. Deift, L. C. Li, T. Nanda and C. Tomei, ``The Toda Flow on
a Generic Orbit is Integrable", Commun. on Pure and Appl. Math., Vol.XXXIX
(1986) 183-232.

\bibitem {Babelon}O. Babelon, C.-M. Viallet, ``Integrable models, Yang-Baxter
equations and quantum groups", Part I, Ref. SISSA 54 EP (1989)

\bibitem {DS}V.G.Drinfeld, Sokolov, J.Sov.Math.Phys. 30 (1984) 1975.

\bibitem {Ueno}K.Ueno, K.Takasaki, ``Toda lattice hierarchy", Advanced Studies
in Pure and Applied Mathematics 4 (1984) 1-95.
\end{thebibliography}
\end{document}